\documentclass[prl,aps,twocolumn,showpacs]{revtex4}
\begin{document}

\def\bra#1{\left\langle#1\right|}
\def\ket#1{\left|#1\right\rangle}

%\draft

\title{Translation of Quantum Texts}
\author{Randall Espinoza, Tom Imbo and Paul Lopata}
\affiliation{Department of Physics, 845 W. Taylor St., University of Illinois at Chicago, Chicago, IL 60607-7059}

\begin{abstract}
In the companion to this paper, we described a generalization of the deterministic quantum cloning process, called enscription, which utilizes entanglement in order to achieve the ``copying'' of (certain) sets of distinct quantum states which are not orthogonal, called texts. Here we provide a further generalization, called translation, which allows us to completely determine all translatable texts, and which displays an intimate relationship to the mathematical theory of graphs. 

\end{abstract}


\maketitle

Assuming the reader to be familiar with the motivation, terminology, and basic results of \cite{foot1}, we proceed directly to the concept which will be the focus of this paper. 
Consider a quantum $N$-text ${T=\{\ket{\psi_i}\}}$ with language $\cal H$, and a composite system with Hilbert space ${\cal H}\otimes{\cal H}$ containing the $N$ entangled states 
\begin{equation}
\ket{\Omega^T_i(q,\psi_0)}={1\over\sqrt{A_i}}\big(\ket{\psi_i}\otimes\ket{\psi_0}+q\ket{\psi_0}\otimes\ket{\psi_i}\big) \label{on}
\end{equation}
for some fixed unit vector $\ket{\psi_0}$ in $\cal H$ and complex number~$q$, where $A_i=1+|q|^2+2\,{\rm Re}(q)\,|\langle\psi_i|\psi_0\rangle|^2$. We say that $T$ can be {\it q-translated} if there exists a state $\ket{\psi_0}$ in~$\cal H$, an $N$-text $T^{\prime}=\{\ket{\chi_i}\}$ with language $\cal H$, and a unitary transformation $U$ on ${\cal H}\otimes{\cal H}$ such that 
\begin{equation}
U\,\ket{\Omega^T_i(q,\psi_0)}=\ket{\chi_i}\otimes\ket{\psi_i}\label{th}
\end{equation}
for each $1\leq i\leq N$. In this case, $\ket{\psi_0}$ is called a {\it q-tablet} for $T$, $U$ is called the {\it procedure} of the {\it q-translation of $T$~onto $\ket{\psi_0}$}, and $T^{\prime}$ is referred to as the {\it output text}. (Note that a $q$-translation of $T$ provides a \mbox{$q$-translation} of every subtext of $T$.)
We drop the ``$q$'' which labels a translation and tablet --- and/or add the adjectives {\it weakly central, central} and {\it quasi-central} --- in exactly the same manner as for enscription.

We will call a translation of a text $T$ {\it faithful} if the output text $T^{\prime}$ is equivalent (see~\cite{foot1}) to $T$. It should be clear that a text can be $q$-enscribed if and only if it can be faithfully $q$-translated. What may be less obvious is that the restrictions on (general) $0$-translations are just as severe as those on $0$-enscriptions (that is, on cloning): {\it A text $T$ can be $0$-translated if and only if $T$ is classical}. (This can be found, in different terminology, on pp.~586-587 of~\cite{foot2}.) The main purpose of this paper is to find an analog of this result with the $q=0$ restriction removed. 

The unitary transformation $U$ in (\ref{th}) exists if and only~if 
\begin{equation}
\langle \Omega^T_i(q,\psi_0)|\Omega^T_j(q,\psi_0)\rangle =\langle \chi_i|\chi_j\rangle\langle\psi_i|\psi_j\rangle \label{fo}
\end{equation}
for all $1\leq i<j\leq N$. (Note that (\ref{fo}) is automatically satisfied when $i=j$.) These ${N(N-1)\over 2}$ conditions can be rewritten as
\begin{equation}
z_{ij}+Q\,\langle \psi_i|\psi_0\rangle\langle\psi_0|\psi_j\rangle = \sqrt{B_iB_j}\,y_{ij}z_{ij}\ \ \ , \label{fi}
\end{equation}
where $y_{ij}=\langle \chi_i|\chi_j\rangle$, $z_{ij}=\langle \psi_i|\psi_j\rangle$, $Q={2\,{\rm Re}(q)\over 1+|q|^2}$, and $B_i=1+Q\,|\langle\psi_i|\psi_0\rangle|^2$. The $N\times N$ matrices defined by $z_{ij}$ and $y_{ij}$ are the Gram matrices of $T$ and $T^{\prime}$ respectively. The real number $-1\leq Q\leq 1$ is called the {\it entanglement parameter} of the translation. We see from (\ref{fi}) that if the complex numbers $q_1$ and $q_2$ lead to the same value of $Q$, then a given text $T$ can be $q_1$-translated if and only if it can be $q_2$-translated. 

The Gram matrix of the output text $T^{\prime}$ of a translation is ``almost'' completely determined by the entanglement parameter $Q$, the tablet $\ket{\psi_0}$, and the ``input text'' $T$ (that is, by~(\ref{fi})). The only matrix elements $y_{ij}$ which are not determined by this data are those where the corresponding matrix elements $z_{ij}$ are zero. Let us call the set ${\cal J}_T$ of all pairs $(i,j)$ such that $z_{ij}=0$ the {\it null index set} of~$T$. We then see that a text $T$ is translatable if and only if there exists a $-1\leq Q\leq 1$, a state $\ket{\psi_0}$ in $\cal H$, and complex numbers $y_{ij}={\overline y_{ji}}$ for all $(i,j)$ in ${\cal J}_T$, such that the matrix defined by these $y$'s and 
\begin{equation}
y_{k\ell}= \frac{1 + Q\,\langle\psi_k|\psi_0\rangle\langle\psi_0|\psi_{\ell}\rangle/z_{k\ell}}{\sqrt{(1+Q|\langle\psi_0|\psi_k\rangle|^2)(1+Q|\langle\psi_0|\psi_{\ell}\rangle|^2)}}\label{out}
\end{equation}
for $(k,\ell )$ not in ${\cal J}_T$, is positive semi-definite with all off-diagonal elements having modulus $<1$. (It can be shown that such a matrix is always the Gram matrix of some text $T^{\prime}$.) In particular, for a fully-quantum text $T$ we have ${\cal J}_T=\emptyset$, so that (given $Q$ and $\ket{\psi_0}$) the Gram matrix of $T^{\prime}$ is completely determined by~(\ref{out}) in this case.

It is straightforward to show that the analogs of \mbox{Theorems~1-5,~7~and~10} in~\cite{foot1} also hold for translation. In~particular, and most importantly for our purposes: (i)~{\it All translatable texts are efficient}, and (ii)~{\it If $T$ is a direct sum of a classical subtext and a translatable subtext, then $T$ is translatable}. But enscription and translation also have their differences. For example, the ``if'' part of Lemma~2 in~\cite{foot1} fails for translation. Instead we have

\vskip 5pt
\noindent
{\bf Lemma~A:} Let $\ket{\psi_0}$ be a $q$-tablet for the translatable text $ T=\{\ket{\psi_i}\}$, with ${\rm Re}(q)\neq 0$. Then for any $i\neq j$
\begin{eqnarray*}
({\rm a})\ \ \langle\psi_0|\psi_i\rangle = 0 \; \; \text{ and } \; \; \langle\psi_0|\psi_j\rangle= 0 & \Rightarrow  & \langle\psi_i|\psi_j\rangle = 0 \\
({\rm b})\ \ \langle\psi_0|\psi_i\rangle \neq 0 \; \; \text{ and } \; \; \langle\psi_0|\psi_j\rangle  \neq 0 & \Rightarrow  & \langle\psi_i|\psi_j\rangle \neq 0.
\end{eqnarray*}

\vskip 3pt
\noindent
{\it Proof:} Both (a) and (b) follow easily from (\ref{fi}). {\it Q.E.D.}

\vskip 5pt
\noindent
We can use Lemma~A to obtain a generalization of Theorem~6 in~\cite{foot1}, but first we will need the following definition. A text $T$ is a {\it disjoint union} of the subtexts $T_1=\{\ket{\psi_1},\dots ,\ket{\psi_{N_1}}\}$ and $T_2=\{\ket{\phi_1},\dots ,\ket{\phi_{N_2}}\}$ if $T=\{\ket{\psi_1},\dots ,\ket{\psi_{N_1}},\ket{\phi_1},\dots ,\ket{\phi_{N_2}}\}$. (This is a weakening of the notion of a direct sum of texts.) 

\vskip 5pt
\noindent
{\bf Theorem~A:} Every translatable text $T$ is a disjoint union of a classical subtext and a fully-quantum subtext.

\vskip 3pt
\noindent
{\it Proof:} Let $\ket{\psi_0}$ be a tablet for $T$. Define the subtext $T_1$ of $T$ to be the set of all states in $T$ which are orthogonal to $\ket{\psi_0}$, and the subtext $T_2$ to contain the remaining states in $T$. Clearly $T$ is a disjoint union of $T_1$ and $T_2$, and it follows from Lemma~A~(a) that $T_1$ is classical and from Lemma~A~(b) that $T_2$ is fully-quantum. {\it Q.E.D.}  

\vskip 5pt
\noindent
The following lemma and definition allow us to bring this further.

\vskip 5pt
\noindent
{\bf Lemma~B:} Let $\ket{\psi_0}$ be a tablet for a translatable, fully-quantum $N$-text $T = \{\ket{\psi_i}\}$, with $N\geq 3$. Then we have $\langle\psi_0|\psi_i\rangle \neq 0$ for any $i$.  

\vskip 3pt
\noindent
{\it Proof:} It suffices to consider the case $N=3$. Assume that the tablet is orthogonal to one of the states in $T$, say $\ket{\psi_3}$. Then from (\ref{out}) it is straightforward to show that the non-negativity of the Gram matrix $y_{ij}$ requires $|z_{12}| \geq 1$. {\it Q.E.D.}

\vskip 5pt
\noindent
A fully-quantum subtext $T^{\prime}$ of a text $T$ is said to be {\it maximal} if $T^{\prime}$ is not contained in any larger fully-quantum subtext of $T$. It is straightforward to show that any text which is a disjoint union of a classical subtext and a fully-quantum subtext, is also a disjoint union of a classical subtext and a maximal, fully-quantum subtext. In particular, by Theorem~A, every translatable text can be expressed as the latter disjoint union. We can now prove

\vskip 5pt
\noindent
{\bf Theorem B:} Let the translatable text $T$ be a disjoint union of a classical subtext $T_1$ and a maximal, fully-quantum subtext $T_2$. Then each state in $T_1$ has a non-zero inner-product with {\it at most} one state in~$T_2$.

\vskip 3pt
\noindent 
{\it Proof:} The maximality of $T_2$ implies that for any fixed $\ket{\psi}$ in $T_1$, there exists a $\ket{\phi}$ in $T_2$ such that $\langle\psi|\phi\rangle= 0$. Now assume that there are two other states $\ket{\phi_1}$ and $\ket{\phi_2}$ in $T_2$ such that $\langle\psi|\phi_1\rangle\langle\psi|\phi_2\rangle\neq 0$. By Lemma~B we have $\langle\phi_1|\psi_0\rangle\langle\phi_2|\psi_0\rangle\langle\phi|\psi_0\rangle\neq 0$, where $\ket{\psi_0}$ is any tablet for $T$. Applying Lemma~B to the fully-quantum subtext $\{\ket{\psi},\ket{\phi_1},\ket{\phi_2}\}$ we similarly obtain $\langle\psi|\psi_0\rangle\neq 0$. But, by Lemma~A~(b), we cannot have $\langle\psi|\phi\rangle =0$ and $\langle\phi|\psi_0\rangle\langle\psi|\psi_0\rangle\neq 0$. {\it Q.E.D.}

\vskip 5pt
\noindent
(The analogous result for enscription would replace ``at most one'' by ``at most zero''. See Theorem~6 in~\cite{foot1}.)

Are the conditions in Theorem~B on the structure of $T$ also {\it sufficient} for the translatability of an efficient text? To answer this, it will be useful to associate with each text $T$ a simple graph $\Gamma (T)$. (A {\it simple graph} consists of a set of ``vertices'', along with a set of ``edges'' connecting pairs of {\it distinct} vertices, each such pair being joined by {\it at most} one edge. In what follows, all graphs are simple.) We construct $\Gamma (T)$ by drawing a vertex for each state in~$T$, and then inserting an edge between two distinct vertices if the corresponding states are not orthogonal. (It can be shown that given any graph~$\Gamma_0$, there exists an efficient text $T$ such that $\Gamma (T)$ is ``isomorphic'' to~$\Gamma_0$.) The results from the previous section can be nicely re-stated in terms of these graphs. But first we will need some terminology (see, for example,~\cite{foot3}). A graph ${\Gamma}^{\prime}$ is an {\it induced subgraph} of a graph $\Gamma$ if (i) the vertex set of ${\Gamma}^{\prime}$ is a subset of the vertex set of $\Gamma$, and (ii)~there is an edge between two vertices of ${\Gamma}^{\prime}$ if and only if the corresponding edge exists in $\Gamma$. (As an example we have that if $T^{\prime}$ is a subtext of $T$, then $\Gamma (T^{\prime})$ is an induced subgraph of $\Gamma (T)$.) A graph is {\it complete} if it has an edge between every pair of vertices, while it is {\it independent} if it has no edges at all. A complete subgraph of a graph~$\Gamma$ (which is necessarily induced) is {\it maximal} if it is not contained in any larger complete subgraph of $\Gamma$. Finally, a graph $\Gamma$ is {\it split} if it has two induced subgraphs $\Gamma_1$ and $\Gamma_2$ which have no vertices in common, but which together exhaust all the vertices of~$\Gamma$, with $\Gamma_1$ independent and $\Gamma_2$ complete. We call such a ``decomposition'' of $\Gamma$ a {\it splitting}. (A simple class of split graphs are the {\it star graphs}, whose only edges emanate from a single vertex and connect to every other vertex in the graph.) An alternative characterization of split graphs is given by the following ``forbidden subgraph'' theorem (see Theorem~7.1.1 in~\cite{foot3}): {\it A graph is split if and only if it contains no induced subgraphs of the type shown in Fig.~1~(a),~(b)~and~(d).}

\begin{figure}
\label{exclude}
\setlength{\unitlength}{1mm}
\begin{picture}(80, 40) 
   \thicklines
   \put(5,27){\circle*{2}}
   \put(5,37){\circle*{2}}
   \put(15,27){\circle*{2}}
   \put(15,37){\circle*{2}}
   \put(5,27){\line(1,0){10}}
   \put(5,37){\line(1,0){10}}	
   \put(8,22){\text{(a)}}
   \put(35,27){\circle*{2}}
   \put(35,37){\circle*{2}}
   \put(45,27){\circle*{2}}
   \put(45,37){\circle*{2}}
   \put(35,37){\line(1,0){10}}
   \put(35,27){\line(1,0){10}}
   \put(35,27){\line(0,1){10}}
   \put(45,27){\line(0,1){10}}
   \put(38,22){\text{(b)}}
  \put(65,27){\circle*{2}}
   \put(65,37){\circle*{2}}
   \put(75,27){\circle*{2}}
   \put(75,37){\circle*{2}}
   \put(65,37){\line(1,0){10}}
   \put(65,27){\line(1,0){10}}
   \put(65,27){\line(0,1){10}}
   \put(75,27){\line(0,1){10}}
   \put(65,27){\line(1,1){10}}
   \put(68,22){\text{(c)}}
   \put(34,1){\circle*{2}}
   \put(46,1){\circle*{2}}
   \put(30,11){\circle*{2}}
   \put(50,11){\circle*{2}}
   \put(40,17){\circle*{2}}
   \put(34,1){\line(1,0){12}}
   \put(34,1){\line(-2,5){4}}
   \put(50,11){\line(-2,-5){4}}
   \put(50,11){\line(-5,3){10}}
   \put(40,17){\line(-5,-3){10}}
   \put(38,-4){\text{(d)}}
\end{picture}
\vskip 3pt
\caption{The forbidden induced subgraphs characterizing split [(a), (b), and (d)] and well-split [(a), (b), (c) and (d)] graphs.}
\end{figure}
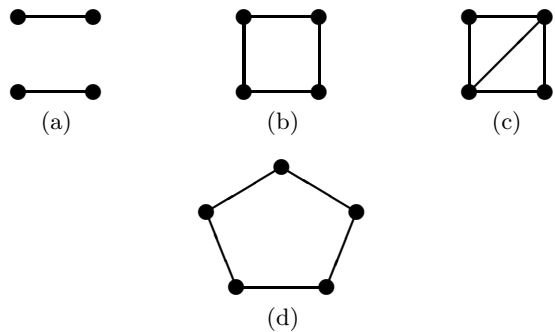

Given a translatable text $T$, the associated graph $\Gamma (T)$ is split (by Theorem~A). That is, if we express $T$ as a disjoint union of a classical subtext $T_1$ and a maximal, fully-quantum subtext $T_2$, there will be a corresponding splitting of $\Gamma (T)$ into the independent subgraph $\Gamma (T_1)$ and the maximal, complete subgraph $\Gamma (T_2)$. Theorem~B then tells us that there is a maximum of one edge connected to each vertex in $\Gamma (T_1)$. In general, we say that a split graph $\Gamma$ is {\it well-split} if for any splitting of $\Gamma$ into an independent induced subgraph $\Gamma_1$ and a maximal, complete subgraph $\Gamma_2$ (a splitting which always exists), there is at most one edge connected to each vertex in $\Gamma_1$. Thus, Theorems~A~and~B can be restated~as

\vskip 5pt
\noindent
{\bf Theorem~C:} Let $T$ be a translatable text. Then $\Gamma (T)$ is well-split.

\vskip 5pt
\noindent
As for split graphs, we can also characterize well-split graphs by a forbidden subgraph theorem.

\vskip 5pt
\noindent
{\bf Theorem~D:} A graph $\Gamma$ is well-split if and only if it has no induced subgraphs of the four types shown in Fig.~1.

\vskip 3pt
\noindent
{\it Proof:} (i) We first treat the ``only if''. Assume $\Gamma$ is well-split. We know that induced subgraphs of the types shown in Fig.~1~(a),~(b),~and~(d) are forbidden since $\Gamma$ is split. The remaining type in Fig.~1 is ruled out by an argument analogous to that in the proof of Theorem~B. (ii)~Now the ``if''. Assume that $\Gamma$ has no induced subgraphs of the types shown in Fig.~1. Clearly $\Gamma$ is split, and hence its vertices can be partitioned into an ``independent set'' $V_1$ and a ``maximal, complete set'' $V_2$. This implies that for every vertex in $V_1$ there exists at least one vertex in $V_2$ to which it is not connected by an edge. Now assume that $\Gamma$ is {\it not} well-split. Then there is some $v$ in $V_1$ which is connected by an edge to (at least) two vertices $w_1$ and $w_2$ in $V_2$. If $w$ denotes a vertex in $V_2$ which is not connected to $v$ by an edge, then the subgraph induced by the vertex set $\{v,w,w_1,w_2\}$ is isomorphic to the graph in Fig.~1~(c). {\it Q.E.D.}

\vskip 5pt
\noindent
In particular, all graphs with less than four vertices are well-split.

It will be convenient to have a parameterization of a generic well-split graph $\Gamma$. We will assume that $\Gamma$ is ``connected'' --- that is, any two vertices in $\Gamma$ can be connected by a ``path'' of edges in $\Gamma$. The generalization to the disconnected case is obvious. (If a non-independent split graph is disconnected, then all but one of its connected components consist of a single vertex.) First, partition the set $V_{\Gamma}$ of vertices of $\Gamma$ as ${V_{\Gamma}=V_1\cup V_2}$, where the subgraph of $\Gamma$ induced by $V_1$ is independent, while that induced by $V_2$ is maximally complete. (This splitting is unique, except in the case where $\Gamma$ is a star graph with three or more vertices.) Let $n_{1(2)}$ denote the number of vertices contained in $V_{1(2)}$. Also, let $\ell\leq n_2$ be the number of vertices in $V_2$ which are connected by an edge to at least one vertex in $V_1$. Finally, define $m_j$ to be the number of vertices in $V_1$ which are connected to the ``$j$-th vertex'' $w_j$ in $V_2=\{w_k\}$. Clearly, $n_1=\sum_jm_j$. It should also be clear that the integers $n_2$, $\ell$, and $m_1,\dots ,m_{\ell}$ completely characterize the isomorphism class of $\Gamma$. We now label each vertex in $V_1$ as $v_{ij}$, where $1\leq i\leq \ell$  represents the ``position'' of the unique vertex $w_i$ in $V_2$ to which it is connected by an edge.  The $j$ index of $v_{ij}$ ($1 \leq j \leq m_i$) then orders the vertices in $V_1$ which are connected to $w_i$. (An example of a well-split graph, with vertices labeled as above, is given in Fig.~2.) Using this parameterization, we prove the following result which shows that (beyond Theorem~C) there are no additional graphical obstructions to translation.

\begin{figure}
\label{well-split}
\setlength{\unitlength}{1mm}
\begin{picture}(40,18) 
   \thicklines
   \put(30,7){\circle*{2}}
   \put(20,1){\circle*{2}}
   \put(20,13){\circle*{2}}
   \put(10,1){\circle*{2}}
   \put(10,7){\circle*{2}}
   \put(10,13){\circle*{2}}
   \put(20,1){\line(5,3){10}}
   \put(20,1){\line(0,1){12}}
   \put(20,1){\line(-5,3){10}} 
   \put(20,1){\line(-1,0){10}}
   \put(20,13){\line(5,-3){10}}
   \put(20,13){\line(-1,0){10}}
   \put(31,8){$w_3$}
   \put(21,-2){$w_2$}
   \put(21,14){$w_1$}
   \put(5,2){$v_{22}$}
   \put(5,8){$v_{21}$}
   \put(5,14){$v_{11}$}
\end{picture}
\vskip 1pt
\caption{An example of a well-split graph with six vertices.} 
\end{figure}
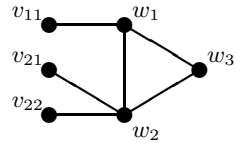

\vskip 5pt
\noindent
{\bf Theorem~E:} Let $\Gamma_0$ be a well-split graph. Then there exists a translatable text $T$ such that $\Gamma (T)$ is isomorphic to $\Gamma_0$.

\vskip 3pt
\noindent
{\it Proof:} It suffices to consider the case where $\Gamma_0$ is connected, since it is easy to extend the proof to include additional isolated vertices. Let $n_2$, $\ell$, and $m_1,\dots ,m_{\ell}$ be the integers (as in the above discussion) which characterize the isomorphism class of $\Gamma_0$, and consider the text $T=\{\ket{\psi_{ij}},\ket{\phi_k}\}$, $1\leq i\leq\ell$, $1\leq j\leq m_{i}$, $1\leq k\leq n_2$, defined by $\langle\phi_k|\phi_{k^{\prime}}\rangle=z$ (for all $k\neq k^{\prime}$), $\langle\psi_{ij}|\psi_{i^{\prime}j^{\prime}}\rangle =0$ [for all $(i,j)\neq (i^{\prime},j^{\prime})$] and $\langle\psi_{ij}|\phi_k\rangle =z_i\,\delta_{ik}$ (for all $i,j,k$), where $-\frac{1}{ n_2-1}<z<0$ and the $z_i$'s are non-zero real numbers. For any such $z$, it is possible to choose the $z_i$'s ``small enough'' so that this defines a positive definite Gram matrix, and hence $T$ is a text (as advertised). By construction, $\Gamma (T)$ is isomorphic to $\Gamma_0$. The null index set ${\cal J}_T$ of $T$ contains the ``pairs'' $[(i,j),(i^{\prime},j^{\prime})]$ for $(i,j)\neq (i^{\prime},j^{\prime})$, and $[(i,j),k]$ for $i\neq k$. Thus, in any translation of $T$, it is these elements of the output Gram matrix which are not determined by~(\ref{out}). Now consider  $T^{\prime}=\{\ket{\chi_{ij}},\ket{\,\eta_k}\}$, $1\leq i\leq\ell$, $1\leq j\leq m_{i}$, $1\leq k\leq n_2$, where $\langle\eta_k|\,\eta_{k^{\prime}}\rangle= \delta_{kk'}$, $\langle\chi_{ij}|\chi_{i^{\prime}j^{\prime}}\rangle =0$ (for all $j,j^{\prime}$, and $i\neq i^{\prime}$), $\langle\chi_{ij}|\chi_{ij^{\prime}}\rangle =1/\sqrt{1-z}$ (for all $i$, and $j\neq j^{\prime}$), $\langle\chi_{ij}|\,\eta_i\rangle =1/\sqrt{1-z}$ (for all $i,j$), and $\langle\chi_{ij}|\,\eta_k\rangle =0$ (for all $j$ and $i\neq k$). For any $z$ as above, we have that $T^{\prime}$ is also a text. Moreover, it is possible to choose $z$ and $0<Q\leq 1$ so that there exists a state $\ket{\psi_0}$ satisfying  $z=-Q\,|\langle\phi_i|\psi_0\rangle|^2$ (for all $i$), and (by judicious choices of the $z_i$'s) such that $\langle\psi_{ij}|\psi_0\rangle=0$ [for all $(i,j)$]. It is then straightforward to check from~(\ref{fi}) that $T$ can be translated onto $\ket{\psi_0}$ with entanglement parameter $Q$ and output text $T^{\prime}$. {\it Q.E.D.}

\vskip 5pt
\noindent
Note that the graph $\Gamma (T^{\prime})$ of the output text in the proof of Theorem~E does not inherit the well-split structure possessed by $\Gamma (T)$ --- the former graph is disconnected, each of the $n_2$ connected components being complete. (Thus, the output text of a translation need not be translatable!) Indeed, we have yet to find any simple relationship between the properties of an input text and those of its possible output texts. As examples of the variety of possibilities, we have \mbox{(i)~every} text is the output text of a translation of a classical text, and (ii)~there exist fully-quantum texts which can be translated with classical output texts. Several techniques designed to address this issue will be presented in~\cite{foot4}.

One might hope for a stronger version of Theorems~C~and~E --- namely, that an efficient text $T$ is translatable {\it if and only if}  $\Gamma (T)$ is well-split. But this is not true. Indeed, there exist efficient texts $T$ with $\Gamma (T)$ complete that cannot be $q$-translated for any $q$, as is shown by the following theorem which determines all translatable fully-quantum texts:

\vskip 5pt
\noindent
{\bf Theorem~F:} Let $T$ be an efficient, fully-quantum text with Gram matrix $z_{ij}$. Then $T$ is translatable if and only if all of the non-zero eigenvalues of the matrix $M_{ij}={1\over z_{ij}}$ have the same sign ``$\epsilon$'' except for one (simple) eigenvalue. ($M$ always has at least one positive and at least one negative eigenvalue.) Moreover, ${{\rm sign}(Q)=\epsilon}$ for any translation of $T$, and $|Q|$ can be chosen arbitrarily small. Finally, there exists an {\it efficient} output text of a translation of $T$ if and only if ${\rm det}\,(M)\neq 0$.

\vskip 5pt
\noindent
(The lengthy proof of this remarkable theorem, which will be given in~\cite{foot4}, highlights intriguing connections between translation and the ``Hadamard product'' of matrices.)
Theorem~9 in~\cite{foot1} is a simple corollary of this result. Also, in contrast to Theorems~8~and~12 in~\cite{foot1}, Theorem~F implies that all efficient real, uniform texts are translatable. (Indeed, they can all be {\it centrally} translated.) Another consequence is

\vskip 5pt
\noindent
{\bf Theorem~G:} All efficient 3-texts can be translated.

\vskip 3pt
\noindent
{\it Proof:}
Let $T$ be an efficient 3-text. If $T$ is fully-quantum, then it is translatable by Theorem~F. If not, then there are either exactly one or exactly two non-zero $z_{ij}$'s ($i<j$) in the Gram matrix of $T$. In the former case, where $\Gamma (T)$ is disconnected, it follows from Theorems~4~and~7 in~\cite{foot1} that $T$ is enscribable, and hence translatable. In the latter case, $T$ is equivalent to a text that can be translated as in the proof of Theorem~E. {\it Q.E.D.}

\vskip 5pt
\noindent
In particular, the first examples of efficient, fully-quantum $N$-texts that are ``untranslatable'' occur at $N=4$, when $M$ has two positive and two negative eigenvalues.

Can we find an analog of Theorem~F beyond the fully-quantum case? As a first step we have

\vskip 5pt
\noindent
{\bf Theorem~H:} Let $T$ be a translatable text which is not fully-quantum, and such that $\Gamma (T)$ is connected. Then every translation of $T$ has $Q>0$.

\vskip 3pt
\noindent
{\it Proof:} If (say) $z_{ij}=0$, we see from the contrapositive of Lemma~A~(b) that either ${\langle\psi_0|\psi_i\rangle =0}$ or $\langle\psi_0|\psi_j\rangle =0$. Assume the former is true. Since $\Gamma (T)$ is connected, we have that $z_{ik}\neq 0$ for some $k\neq i$. Then if $Q\leq 0$, we have from~(\ref{out}) that $|y_{ik}|\geq 1$. {\it Q.E.D.}

\vskip 5pt
\noindent 
Interestingly, this gives (beyond Theorem~F) the only additional non-graphical obstruction to translation.

\vskip 5pt
\noindent
{\bf Theorem~I:} Let the text $T$, with $\Gamma (T)$ connected and well-split, be a disjoint union of a (non-empty) classical subtext~$T_1$ and a maximal, fully-quantum subtext $T_2$. Then $T$ is translatable if and only if $T_2$ is translatable with $Q>0$.

\vskip 3pt
\noindent
{\it Proof:} The ``only if'' follows immediately from Theorem~H. For the converse, first consider the case where $T_1$ contains just one state $\ket{\phi}$, whose only non-zero inner product is with (say) $\ket{\psi_1}$ in $T_2=\{\ket{\psi_i}\}$. Let $\ket{\eta_0}$ be a tablet for $T_2$ associated with the entanglement parameter $Q_2>0$ and the output text $T_2^{\prime}=\{\ket{\chi_i}\}$. (Without loss of generality, we may assume that $\ket{\eta_0}$ lives in the dialect ${\cal H}_{T_2}$ of~$T_2$.)  Next, define $T^{\prime}=\{\ket{\chi_i},\ket{\chi}\}$ where $$\ket{\chi}=\frac{1}{\sqrt{B_1}}\ket{\chi_1}+\sqrt{\frac{B_1-1}{B_1}}\ket{\chi_{\perp}},$$ and the state $\ket{\chi_{\perp}}$ is orthogonal to every state in $T_2^{\prime}$. Finally, let $${\ket{\psi_0}=\alpha\ket{\eta_0}+\beta (I-P)\ket{\phi}}\ ,$$ where $I$ is the identity operator on $\cal H$ (the language of~$T$), $P$ is the projection operator onto ${\cal H}_{T_2}$, and
$$|\alpha |^2=1-|\beta |^2|\langle\phi|(I-P)|\phi\rangle |^2={1\over 1+|\langle\phi|\eta_0\rangle |^2}\ .$$ (We fix the relative phase of $\alpha$ and $\beta$ by requiring $\langle\phi|\psi_0\rangle =0$.) It is then straightforward to show, using~(\ref{fi}), that $T$ can be translated onto $\ket{\psi_0}$ with output text $T^{\prime}$ and entanglement parameter $Q={1\over |\alpha|^2}Q_2$ (as long as $Q_2$ is chosen small enough so that $Q\leq 1$, which is always possible by Theorem~F). If $T_1$ contains more than one state, we use the above procedure iteratively (along with the fact that the entanglement parameter at any stage can be made arbitrarily small) to show that $T$ is translatable. {\it Q.E.D.} 

\vskip 5pt
\noindent
We now have a complete classification of translatable texts, given by Theorems~C,~F,~and~I, along with the results (i) and (ii) stated prior to Lemma~A. (\mbox{Theorem~E} can now be viewed as a corollary of this classification.) Numerous other results, techniques, and applications concerning enscription and translation will be developed in~\cite{foot4}.

%requirement of efficiency and the fact that if $T$ is a direct sum of a classical text and a translatable text, then $T$ is %translatable (which we noted prior to Lemma~A).
Since this paper is appropriately entangled with its companion, we can enscribe the acknowledgements from the latter here:
We thank Klaus Bering and Mark Mueller for important contributions. This work was supported in part by the U.S. Department of Energy.

\end{document}